\documentclass[pra,twocolumn,floatix,amssymb,showpacs,amsmath,superscriptaddress]{revtex4-1}
\usepackage{graphicx}
\usepackage{epsfig,color}
\usepackage{amsmath,amssymb,amsthm}
\usepackage{amsmath,amssymb,amsfonts}
\usepackage{algorithmic}
\usepackage{graphicx}
\usepackage{textcomp}
\usepackage{xcolor}

\begin{document}

\title{Challenges in Open-air Microwave Quantum Communication and Sensing}

\author{Mikel Sanz}
\affiliation{Department of Physical Chemistry, University of the Basque Country UPV/EHU, Apartado 644, E-48080 Bilbao, Spain}
\email{mikel.sanz@ehu.eus}
\author{Kirill G. Fedorov}
\affiliation{Walther-Mei{\ss}ner-Institut, Bayerische Akademie der Wissenschaften, D-85748 Garching, Germany}
\affiliation{Physik-Department, Technische Universit\"at M\"unchen, D-85748 Garching, Germany}
\author{Frank Deppe}
\affiliation{Walther-Mei{\ss}ner-Institut, Bayerische Akademie der Wissenschaften, D-85748 Garching, Germany}
\affiliation{Physik-Department, Technische Universit\"at M\"unchen, D-85748 Garching, Germany}
\affiliation{Nanosystems Initiative Munich (NIM), Schellingstraße 4, 80799 M\"unchen, Germany}
\author{Enrique Solano}
\affiliation{Department of Physical Chemistry, University of the Basque Country UPV/EHU, Apartado 644, E-48080 Bilbao, Spain}
\affiliation{IKERBASQUE, Basque Foundation for Science, Maria Diaz de Haro 3, E-48013 Bilbao, Spain}
\affiliation{Department of Physics, Shanghai University, 200444 Shanghai, China}

\begin{abstract}
Quantum communication is a holy grail to achieve secure communication among a set of partners, since it is provably unbreakable by physical laws. Quantum sensing employs quantum entanglement as an extra resource to determine parameters by either using less resources or attaining a precision unachievable in classical protocols. A paradigmatic example is the quantum radar, which allows one to detect an object without being detected oneself, by making use of the additional asset provided by quantum entanglement to reduce the intensity of the signal. In the optical regime, impressive technological advances have been reached in the last years, such as the first quantum communication between ground and satellites, as well as the first proof-of-principle experiments in quantum sensing. The development of microwave quantum technologies turned out, nonetheless, to be more challenging. Here, we will discuss the challenges regarding the use of microwaves for quantum communication and sensing. Based on this analysis, we propose a roadmap to achieve real-life applications in these fields.
\end{abstract}
\maketitle

\section{Introduction}
Quantum communication is a branch of quantum information which makes use of quantum entanglement as a resource to protect and transmit through a quantum channel information with higher capacity than any classical channel and, when combined with quantum cryptography, with provable protection against eavesdropping. In recent years, the experimental progress in this area with optical photons has been astonishing, including a $143$~km quantum communication between the Spanish islands of Tenerife and La Palma \cite{HSFHWUZ15}, a $96$~km connection between Sicily and Malta through a submarine cable \cite{Wetal18}, or the recent quantum communication and quantum key distribution using satellites \cite{Letal17,Retal17,Yetal17,Letal18}, among others. 

Quantum sensing is defined as the use of quantum systems and properties, especially entanglement, as an extra resource to perform a measurement of a physical quantity with higher accuracy or smaller number of classical resources than any classical protocol \cite{DRC17}. Originally, the use of light for quantum sensing was mainly focused on employing continuous-variable quantum states squeezed below the vacuum. In this context, a particularly interesting example is the use of squeezed light for the detection of gravitational waves in LIGO \cite{OIMTBME16}. A modified version of the Hanbury-Brown-Twiss experiment using entangled light allowed for the outperformance the spatial resolution of microscopes below the diffraction limit \cite{SLTIDO13}. A particularly interesting application for our discussion is the application of entangled light to quantum illumination, in which instead of the resolution, the entanglement is used to improve the contrast \cite{Ll08,LRBDOBG13,BGB10}. Finally, it is noteworthy to mention that also impressive advances have been achieved in Heisenberg-limited interferometers by using Fock states \cite{NOOST07}.

\section{State of the art in quantum microwave technology}
The advances in the use of microwaves in the quantum regime for technological applications were more gradual than with optical photons. The reasons are not only historical, but they also lie in technological difficulties which make the control of microwave photons much subtler than optical photons. In this section, we will first address some of the most relevant physical and technological problems of propagating quantum microwaves. Afterwards, we will briefly review the state of the art in experiments and some relevant experimental proposals.

\subsection{Technological Challenges for Quantum Microwaves}

\begin{itemize}
\item The most important challenge when employing microwaves for quantum technologies when compared with optical photons is the requirement of cryogenics. Indeed, the thermal isolation required for photons in the gigahertz regime is much higher than in the terahertz regime. This can be shown by considering the Bose-Einstein distribution, which estimates the number of photons per volume unit with frequency between $\nu$ and $\nu+d\nu$ 
\begin{equation}\label{BE}
n(\nu) \propto \frac{1}{e^{\frac{h \nu}{k_b T}}-1},
\end{equation}
where $T$ is the temperature. If we consider an optical frequency of $\nu=500$ THz at room temperature $T=300$ K, then $n(500\, \text{THz}) \approx 2 \times10^{-35}$ photons. However, if we do the same for $\nu=5$ GHz, we obtain $n(5\, \text{GHz}) \approx 1250$ photons. Therefore, as the energy is much smaller, many thermal microwave photons are created, which is not favorable for quantum applications. This is the fundamental reason why superconducting circuits, which typically operates in the $2-7$ GHz regime use cryogenics at $30$~mK, temperature at which $n(5\, \text{GHz})\approx 3\times10^{-4}$ photons.\
\item The aforementioned fundamental difficulty due to thermal photons seems to limit possible applications of quantum microwaves to intra-fridge environments. This is sufficient for quantum computing applications with superconducting circuits, but only allows for proof-of-principle experiments in quantum communication and sensing.\
\item Photodetectors are devices which transform photons into an electric current, usually by means of a p-n junction or the photoelectric effect. These effects very well fit with optical frequencies, what allows for the construction of photodetectors and photocounters, for flying photons, i.e. for photons which are not trapped inside a cavity. However, this approach cannot be directly applied to propagating quantum microwaves and only photodetectors with limited efficiency (or photodetectors for trapped photons) have been constructed so far. Nonetheless, for most applications in quantum sensing and especially quantum illumination, efficient photodetection is mandatory \cite{SlHG-RSdC17,lHdCFDSS17}, thus further research in required in this area.\
\item The lack of efficient photodetectors reduces the measurements achievable for propagating quantum microwaves to the quantification of electromagnetic field quadratures. 
\end{itemize}

\subsection{Experimental State of the Art}
Intra-fridge technology related to quantum microwaves has experienced a strong advance in the last decades driven by the development of superconducting circuits and superconducting qubits for quantum simulations and quantum computing. Profound improvements have been performed in the fabrication of purer superconducting thin field materials, in the accuracy of lithography, and in the efficiency of cooling down superconducting circuits with the help of closed-cycle dilution refrigerators.

Important theoretical development and experimental achievements towards quantum teleportation with discrete degrees of freedom \cite{Ketal18} and continuous variables \cite{Detal15,Fetal16,Fetal18} in the microwave regime have appeared in the last years. However, these experiments are currently still limited to the intra-fridge environments or work in progress. 

\begin{figure}[t]
\includegraphics[width=0.50\textwidth]{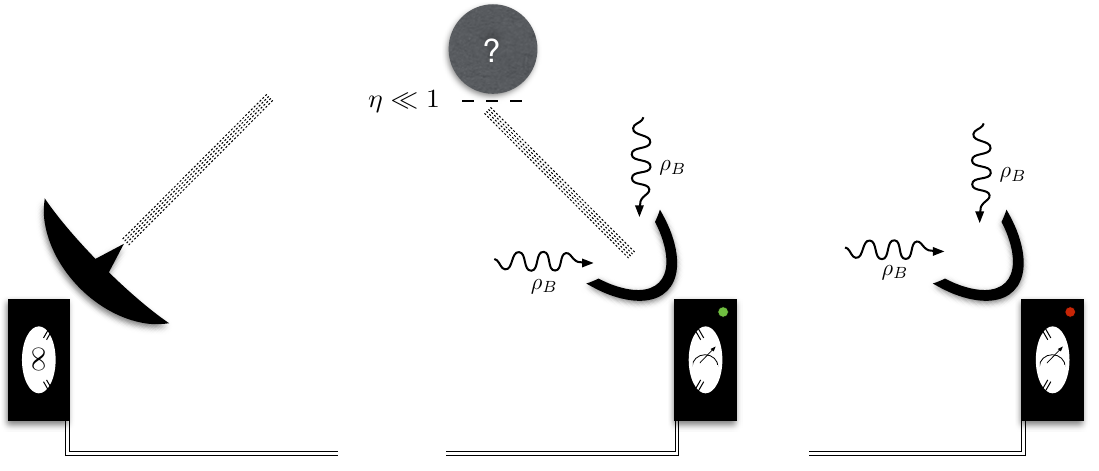}
\caption{A quantum radar makes use of quantum entanglement to enhance the accuracy of a radar or to reduce the use of photons to detect the presence of a low-reflectivity object with reflectivity, $\eta \ll 1$, in a noisy thermal environment $\rho_B$. In the lab, two entangled beams $|\Psi_{SI}\rangle$ are prepared, sending one to the position in which the object might be, while keeping the other in the lab. Afterwards, a joint measurement of the reflected signal together with the beam in the lab is performed. In comparison to classical light, the existence of quantum correlations allows us to achieve up to $6$ dB of quantum advantage in terms of signal-to-noise ratio.}
\label{fig2}
\end{figure}

In the case of quantum illumination or quantum metrology employing quantum microwaves, there are no experimental results to the best of our knowledge. In quantum illumination, entangled radiation could be employed to enhance the detection accuracy of a radar or to reduce the amount of photons demanded to detect the presence of a low-reflectivity object in a noisy environment (detect it without being detected). In general, the idea consists in preparing a pair of entangled optical or microwave beams and irradiating the target with one of them, while preserving the other one in the lab. In comparison to classical light, the existence of quantum correlations between the two beams allows us to declare the presence or absence of the object with either a higher accuracy or less resources, theoretically achieving up to $6$ dB of quantum advantage in terms of signal-to-noise ratio. From the theoretical point of view, the first application of quantum microwaves to quantum illumination was recently introduced in Ref.~\cite{BGWVSP15}. However, this article avoids the problem of the microwave photodetector by using a microwave-to-optics transducer and employing optical photodetectors. Unfortunately, such a transducer has turned out to be as technologically demanding as photodetection in microwaves, and the efficiency of current implementations is not at all sufficient for practical applications. Other theory proposals in the microwave regime aim at the detection of cloaked objects \cite{lHdCFDSS17} or use quantum estimation techniques to obtain the optimal observables to measure \cite{SlHG-RSdC17}.

The lack of single-photon photodetectors for propagating microwaves is one of the main challenges for any possible application of quantum microwaves in quantum communication and sensing. Physically, the reason of the difficulty in developing efficient microwave photodetectors is that a microwave photon energy is four orders of magnitude smaller than the energy of its optical counterpart. Consequently, triggering out a photocurrent is obviously much more difficult in the microwave regime. Additionally, traditional applications of propagating quantum microwaves do not make use of photodetection. For this reason, even though there are proposal for photodetectors from a decade ago \cite{MHH06,RG-RS09,Cetal11,PRJWSG-R11,SSJ16,KKTNN18}, there are no experiments yet for propagating photons not already trapped inside a cavity. To the best of our knowledge, experimental reports deal with "gated" microwave photodetectors, where the time window or even the envelope of the incoming photon is assumed to be known. 

\subsection{Advantages of Microwaves}
We have previously exposed the challenges when using quantum microwaves for quantum communication and sensing, but there are also some important advantages which might be worth the effort. Let us now summarize them:
\begin{figure}[t]
\includegraphics[width=0.50\textwidth]{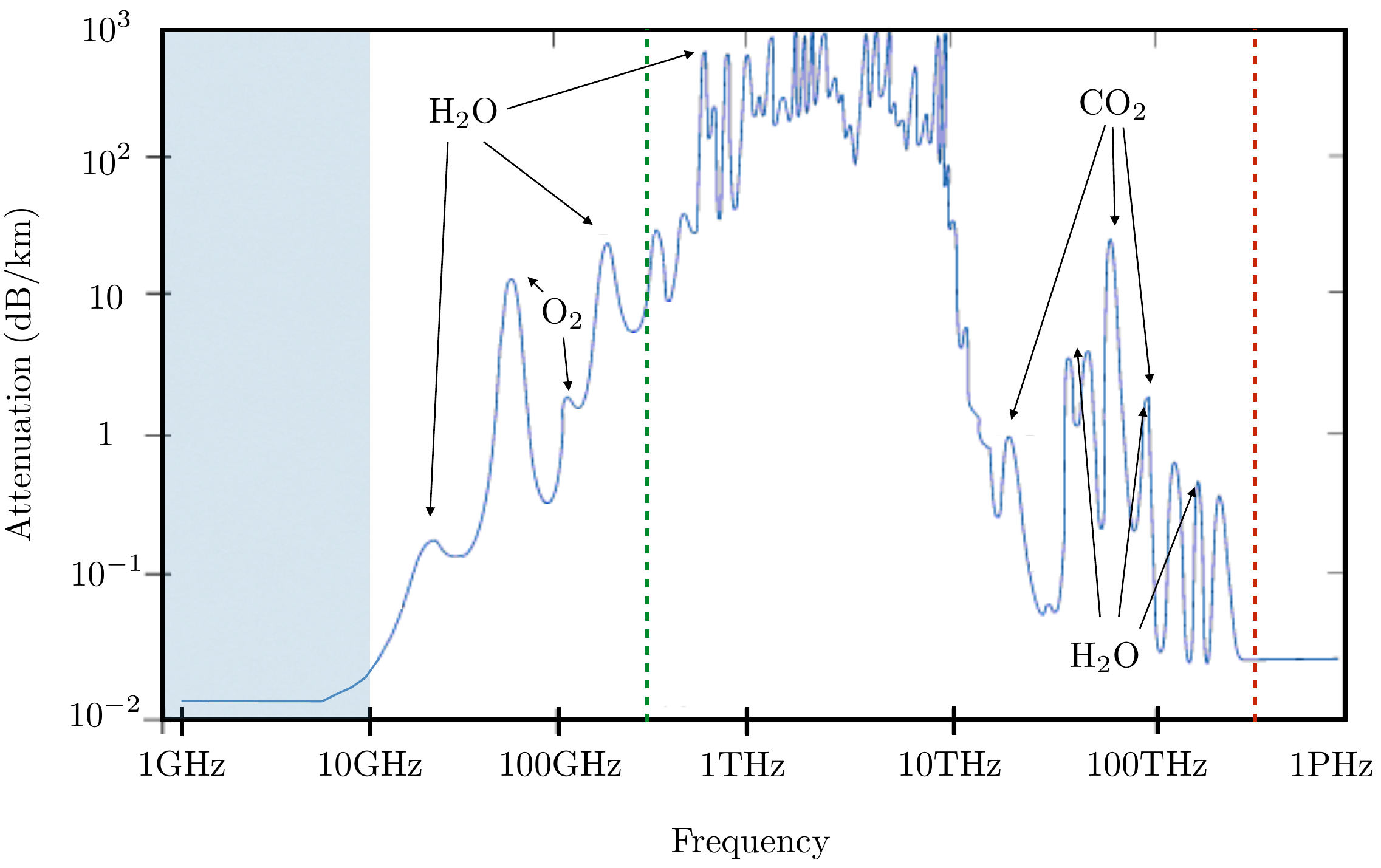}
\caption{Atmospheric attenuation (dB/km) as a function of the frequency (at $20$¼C, $1$ atm and $7.5$ g/m${}^3$ of water). The red dashed line at right separates visible frequencies from infrared, while the green dashed line separates infrared from microwaves. The blue area corresponds to the frequencies in which the technology of propagating quantum microwaves and superconducting circuits operates. This plot uses data taken from Ref.~\cite{Le95}.}
\label{fig1}
\end{figure}
\begin{itemize}
\item Currently, the most advanced and promising (actively developed by big companies such as Google, IBM, Intel, etc.) platform for future quantum computing and simulations is superconducting circuits. These typically consist of thin film layers of superconducting materials and Josephson junctions comprising altogether a network of resonators and qubits. They interact precisely by means of microwave photons, and the quantum state of the qubits is read out by measuring these photons. Quantum information processing occurs inside a fridge at milikelvin temperatures, since higher temperatures imply additional thermal photons which destroy coherence. Controllability and connectivity, as well as a future emergence of a quantum internet, would suggest that distributed quantum computing with a local network of quantum processors must be considered~\cite{Ca17, CCB18}. The use of microwave-optics-microwave transduction has been suggested but, as aforementioned, it is still not sufficiently efficient to date. Consequently, a direct quantum communication with microwaves constitutes a desirable goal.\
\item The smaller energy associated with microwave photons, which is a downside in photodetection, turns into an important advantage when energy consumption is taken into consideration.\
\item Atmosphere frequency-dependent losses contain two low-opacity windows, one in the visible spectrum, and one with even lower attenuation in the frequency range of $100$ MHz-$10$ GHz, as depicted in Fig.~\ref{fig1}. Especially under rainy conditions, microwaves are consequently a suitable frequency range for quantum radar and quantum communication applications.\
\item There are important advantages of using microwaves in quantum radars. There are radars using lower frequencies, but microwaves are convenient for two main reasons: 1) Firstly, objects whose size is comparable to microwave wavelength, such as vehicles, ships, and airplanes, produce large reflexions in this frequency range. The reason is that the range resolution, i.e. the accuracy ascertaining the distance of the target, is determined by the bandwidth of the signal. In this consideration, from the point of view of circuit design, a lower bandwidth is better; 2) Secondly, a narrow beam is usually required, which is afterwards used to scan around to locate the target, but the width of the beam is inversely proportional to the frequency of a given antenna.
\end{itemize}

\section{Roadmap Towards Open-Air Applications}
In this Section, we propose a possible roadmap of theoretical and experimental achievements which must be performed in order to eventually achieve quantum communication and quantum illumination applications of propagating quantum microwaves in the near future. 
\subsection{Intra-Fridge Experiments}
Taking into account the current state of the art in propagating quantum microwaves, the first milestone would be quantum teleportation of continuous-variable quantum states in a fridge. There are two possibilities to attain this task, namely, making use of photodetectors, as usually performed in the optical regime \cite{HSFHWUZ15}, or employing a different approach with only quadrature measurements \cite{Detal15}. Some steps in this direction have already been taken \cite{Fetal16,Fetal18}, but there are still some intermediate stages, associated, for instance, with continuous-variable quantum state transfer. Deterministic state transfer and entanglement protocols between two superconducting qubits fabricated on separate chips have already been achieved in Ref.~\cite{Ketal18}, which is especially interesting for distributed quantum computing. However, the resilience of Fock states, when propagating in a thermal environment, could be smaller than the one corresponding to propagating squeezed Gaussian states.\

Similarly, the first experiments on quantum illumination should be performed in an intra-fridge setting. To our knowledge, no experiments on quantum illumination using microwaves has been realized so far, so we should aim at performing the first proof-of-principle experiments artificially introducing thermal noise and exploring the limits of the quantum signal-to-noise enhancement. Another interesting proof-of-principle experiment would be the use of quantum illumination protocol to detect phase-shift induced cloaking in a highly noisy environment~\cite{lHdCFDSS17}. \

Simultaneously, we must develop single-photon microwave photodetection for flying photons in order to obtain microwave photodetectors with sufficiently high efficiency. A considerable advance in this direction will substantially increase the potential success finding real-life and commercial applications of propagating quantum microwaves. As discussed before, prominent examples here are quantum microwave radars or a microwave quantum local area network (LAN). We highlight two promising approaches for building an efficient microwave photodetector, namely, qubit-based photodetectors~\cite{BGCWKPEW18} and bolometer-based photodetectors~\cite{GLTM16}. Both show advantages and disadvantages, so it would be relevant to follow both approaches in order to determine which one is more advantageous at the end for a given problem.  

\subsection{Transition from Intra- to Inter-Fridge Experiments}
The following natural step is the connection of two different fridges through a millikelvin transmission line. While this may present a challenging engineering task, it is not fundamentally difficult. In contrast, a transmission line at high temperatures ($T=4-300$~K) is a challenging question even from a theory point of view, since universal models to incorporate losses and decoherence of quantum states due to the contact with a room-temperature thermal bath are not at all tested yet. In any case, we expect them to be extremely sensitive to the targeted temperature. The losses can be modeled by a beam-splitter coupling the signal to a frequency-dependent thermal bath. If the temperature is not uniform along the waveguide, we can divide the waveguide into small pieces with constant temperature and reflectivity and analyze the continuous limit.

In the field of quantum illumination, a subsequent key experiment to determine the feasibility of the research line in real-life applications consists in a detection problem in which the object and the source of entangled microwave radiation are in different fridges, connected through a noisy waveguide. One of the beams remains in the original fridge, while the other travels through the highly noisy waveguide, interacts (or not) with a low reflectivity mirror which models the object and, by detecting the reflected signal, one must decide whether the object is there or not. Also for this scenario, the underlying theory work remains to be done.

Finally, quantum microwave technology would strongly profit from engineering filters for thermal photons, the development of non-reciprocal devices which enhance the efficiency of photodetectors, and the development of compact delay lines based, for instance, on Josephson metamaterials.

\subsection{Antennae and Open-Air Challenges}
The emission and reception of microwave quantum signals by means of antennae is still an open problem, since it might require new theoretical and technological developments. Both the theoretical analysis and the type of antenna are dependent on the type of the signal codification (in polarization degrees of freedom, or in Fock basis, etc). However, in both cases of low-aperture antennae, such as parabolic or horn antennae, which can focus the signal in a given direction, seem especially suitable for these purposes. 

The case of entanglement in polarization is somehow simpler, since there are {\it linearly polarized} and {\it circularly polarized} antennae, a feature which is sometimes used to double the information transmission rate by codifying half of the information in one polarization and half in the other. As the antenna polarization matching is required, the information is well preserved. Then, it is natural to consider the option of sending entangled quantum states in polarization degrees of freedom. However, one must take into account that usual technology of superconducting circuits inhibits polarization as a degree of freedom by projecting the electromagnetic field in the waveguide. There are recently relevant advances in 3D superconducting cavities, which could deal also with polarization of the microwave photons, but the technology is by far not as developed as on-chip quasi-1D quantum electrodynamics yet.

For the purposes related to quantum communication, quantum radar and the implementation of other existing protocols, entanglement in the number of photons seems more suitable. More technically, an antenna can be considered as an impedance matcher between the superconducting circuit ($50$ Ohm) and open-air ($377$ Ohm). The paradigmatic setup would consist of a quantum source, for instance, a Josephson parametric amplifier (JPA)\cite{Poetal17}, generating squeezed states, connected by a superconducting waveguide with a $50$ Ohm impedance which ends up in the antenna. The antenna smoothly transforms the impedance of the waveguide into the $377$ Ohm impedance of air. The electromagnetic field propagates throughout the space until the receiving antenna takes care of the opposite matching.

\begin{table}[t]
\caption{Isotropic path losses $L_{\rm P}$ and absorption losses $L_{\rm A}$.}
\begin{center}
    
    \begin{tabular}{|c|c|c|c|c|}
    \hline
            & \multicolumn{2}{|c|}{\parbox{32mm}{\centering\textbf{Optics}   \\$\lambda\,{=}\,810\,\mathrm{nm}$, $\nu\,{=}\,370\,\mathrm{THz}$}} 
						& \multicolumn{2}{|c|}{\parbox{32mm}{\centering\textbf{Microwave}\\$\lambda\,{=}\,60\,\mathrm{mm}$, $\nu\,{=}\,5\,\mathrm{GHz}$ }} \\
    \hline
    d (km)         & $L_{\rm P}$ (dB) & $L_{\rm A}$ (dB)         & $L_{\rm P}$ (dB) &  $L_{\rm A}$ (dB)          \\
    \hline
    $\simeq 1$     & 204       & $3 \cdot 10^{-2}$ & 106       &  $9 \cdot 10^{-3}$  \\
    \hline
    $\simeq 100$   & 244       & $3$             & 146       &  $0.9$              \\
    \hline
    $\simeq 1000$  & 264       & $30$             & 166       &  $9.0$              \\
    \hline
    \multicolumn{4}{l}{}
    \end{tabular}
\label{TabLosses}
\end{center}
\end{table}

Let us briefly analyze communication losses for two typical examples, one in the optical ($\lambda\,{=}\,810\,\textrm{nm}$) and the other in the microwave ($\lambda\,{=}\,60\,\textrm{mm}$) regime. To be as general as possible, we employ the well-known Friis' formula for the calculation of the path-losses
\begin{equation}
\label{Losses}
L = L_{\rm A} G_{\rm t} G_{\rm r} L_{\rm P} = L_{\rm A} G_{\rm t} G_{\rm r} \left( \frac{\lambda}{4 \pi d} \right)^2\,.
\end{equation}
Here, $\lambda$ is the wavelength of the signal, $d$ the distance, $G_{\rm t}$ ($G_{\rm r}$) the gain of the transmitter (receiver), $L_{\rm A}$ are the absorption losses, and $L_{\rm P}$ the free-space path losses for isotropic antennas. Values for $L_{\rm P}$ and  $L_{\rm A}$ are shown for several characteristic distances in Tab.\,\ref{TabLosses}. The results clearly demonstrate that the pure absorption losses are negligible in both frequency regimes for $d\,{\lesssim}\,100\,\textrm{km}$ as it would be the case for links to a satellite where absorption losses practically vanish above a height of ${\simeq}\,10\textrm{km}$. Then, the dominating (isotropic) path losses obviously decrease for smaller frequencies, which is an encouraging result for the application of microwaves.

In order to approach a more realistic situation, we should additionally consider the antenna gains. Assuming a perfect parabolic geometry and zero Ohmic losses, the antenna gain can be estimated via the divergence of the beam it produces. Using this approach, we obtain $G_{\rm t,r}\,{=}\,4\pi/\Omega_{\rm r,t}$, where $\Omega_{\rm r,t}$ are the solid angles associated with the beam divergences $\theta_{\rm t,r}\,{=}\,\lambda/D_{\rm t,r}$ and $D_{\rm t,r}$ are the diameters of the transmitter and receiver antenna apertures. For our optics example~\cite{Yetal17}, antenna gains of $130\,\mathrm{dB}$ can be reached for reasonable aperture diameters $D_{\rm t,r}\,{\simeq}\,1\,\textrm{m}$. The total loss $L$ is therefore reduced to approximately $10\,\textrm{dB}$ for parameters similar to those of the actual satellite quantum communication link demonstrated in Ref.~\cite{Yetal17}. The real losses measured in that work are higher (${\simeq}\,65-82\,\mathrm{dB}$) due to additional experimental imperfections not captured by our idealized model. Nevertheless, this experimental attenuation was small enough to successfully implement quantum key distribution with sifted key rates of a few kb/s~\cite{Letal18}. For microwave frequencies, however, the predicted overall losses are higher. Here, due to the larger wavelength, the decreasing antenna gains outweigh the improvement in isotropic path loss. Increasing the antenna gain via a larger aperture could be useful in quantum illumination (radar) type applications. In quantum communication scenarios, where large antennas may not be very practical, the parameters from the optics experiment~\cite{Yetal17,Letal18} in principle still allow one to cover distances ${\lesssim}\,100\,\textrm{km}$. Naturally, the design of heavily optimized microwave antennas will be mandatory to reach this challenging goal. For communication distances of a few kilometers, the respective requirements are already considerably relaxed and can be viewed as realistic for the near future.

In summary, despite some challenges quantum microwaves propagating in and open air environment exhibit a considerable potential for future applications in quantum communication and sensing. To this end, the development and investigation of suitable antennas and the development of suitable theory models will be key tasks.
\section*{Acknowledgment}

The Authors thank M. M\"ott\"onen, B. Huard, and Y. Omar for the useful discussions. Authors acknowledge the EU Flagship project QMiCS. M. S. and E. S. are grateful for the funding of Spanish MINECO/FEDER FIS2015-69983-P and Basque Government IT986-16. This material is also based upon work supported by the U.S. Department of Energy, Office of Science, Office of Advance Scientific Computing Research (ASCR), under field work proposal number ERKJ335. K. G. F. and F. D. additionally acknowledge financial support from the German Research Foundation through FE 1564/1-1, the doctorate program ExQM of the Elite Network of Bavaria, and the International Max Planck Research School Quantum Science and Technology.


\begin{thebibliography}{00}
\bibitem{HSFHWUZ15} 
T. Herbst, T. Scheidl, M. Fink, J. Handsteiner, B. Wittmann, R. Ursin, and A. Zeilinger, ``Teleportation of entanglement over 143 kmÓ, Proceedings of the National Academy of Sciences, p. 201517007 (2015). 
\bibitem{Wetal18} 
S. Wengerowsky {\it et al.}, ``In-field entanglement distribution over a 96 km-long submarine optical fibre", arXiv preprint:1803.00583 (2018).
\bibitem{Letal17} 
S.-K. Liao {\it et al.}, ``Satellite-to-ground quantum key distribution", Nature, vol. 549, p. 43 (2017). 
\bibitem{Retal17} 
J.-G. Ren {\it et al.}, ``Ground-to-satellite quantum teleportation", Nature, vol. 549, p. 70 (2017). 
\bibitem{Yetal17}
J. Yin {\it et al.}, ``Satellite-based entanglement distribution over 1200 kilometers", Science, vol. 356, p. 1140 (2017).
\bibitem{Letal18}
S.-K. Liao {\it et al.}``Satellite-Relayed Intercontinental Quantum Network", Phys. Rev. Lett., vol. 120, p. 030501 (2018).
\bibitem{DRC17} 
C. L. Degen, F. Reinhard, and P. Cappellaro, ``Quantum sensing", Rev. Mod. Phys., vol. 89, p. 035002 (2017).
\bibitem{OIMTBME16}
E. Oelker, T. Isogai, J. Miller, M. Tse, L. Barsotti, N. Mavalvala, and M. Evans,  ``Audio-Band Frequency-Dependent Squeezing for Gravitational-Wave Detectors", Phys. Rev. Lett., vol. 116, p. 041102 (2016).
\bibitem{SLTIDO13}
O. Schwartz, J. M. Levitt, R. Tenne, S. Itzhakov, Z. Deutsch, and D. Oron, ``Superresolution microscopy with quantum emitters", Nano Lett., vol. 13, p. 5832 (2013).
\bibitem{Ll08}
S. Lloyd, ``Enhanced Sensitivity of Photodetection via Quantum Illumination", Science, vol. 321, p. 1463 (2008).
\bibitem{LRBDOBG13}
E. D. Lopaeva, I. Ruo Berchera, I. P. Degiovanni, S. Olivares, G. Brida, and M. Genovese, ``Experimental Realization of Quantum Illumination", Phys. Rev. Lett., vol. 110, p. 153603 (2013).
\bibitem{BGB10}
G. Brida, M. Genovese, and I. R. Berchera, ``Experimental realization of sub-shot-noise quantum imaging", Nat. Photonics, vol. 4, p. 227 (2010).
\bibitem{NOOST07}
T. Nagata, R. Okamoto, J. L. OÕBrien, K. Sasaki, and S. Takeuchi, ``Beating the Standard Quantum Limit with Four-Entangled Photons", Science, vol. 316, p. 726 (2007).
\bibitem{SlHG-RSdC17}
M. Sanz, and U. Las Heras, J. J. Garcia-Ripoll, E. Solano, and R. Di Candia, ``Quantum Estimation Methods for Quantum Illumination", Phys. Rev. Lett., vol. 118, p. 070803 (2017).
\bibitem{lHdCFDSS17}
U. Las Heras, R. Di Candia, K. G. Fedorov, F. Deppe, M. Sanz, and E. Solano, ``Quantum illumination reveals phase-shift inducing cloaking", Scientific Reports, vol. 7, p. 9333 (2017). 
\bibitem{Ketal18}
P. Kurpiers {\it et al.}, ``Deterministic Quantum State Transfer and Generation of Remote Entanglement using Microwave Photons", Nature, vol. 558, p. 264 (2018).
\bibitem{Detal15}
R. Di Candia {\it et al.}, ``Quantum teleportation of propagating quantum microwaves", EPJ Quantum Technology, vol. 2, p. 25 (2015).
\bibitem{Fetal16}
K. G. Fedorov {\it et al.}, ``Displacement of propagating squeezed microwave states", Phys. Rev. Lett., vol. 117, p. 020502 (2016).
\bibitem{Fetal18}
K. G. Fedorov {\it et al.}, ``Finite-time quantum correlations of propagating squeezed microwaves", Scientific Reports, vol. 8, p. 6416 (2018).
\bibitem{BGCWKPEW18}
J.-C. Besse, S. Gasparinetti, M. C. Collodo, T. Walter, P. Kurpiers, M. Pechal, C. Eichler, and A. Wallraff, ``Single-Shot Quantum Non-Demolition Detection of Itinerant Microwave Photons", Phys. Rev. X, vol. 8, p. 021003 (2018).
\bibitem{BGWVSP15}
S. Barzanjeh, S. Guha, C. Weedbrook, D. Vitali, J. H. Shapiro, and S. Pirandola, ``Microwave Quantum Illumination", Phys. Rev. Lett. vol. 114, p. 080503 (2015).
\bibitem{MHH06}
G. N. Milford, C. C. Harb, and E. H. Huntington, ``Shot noise limited, microwave bandwidth photodetector design", Rev. Sci. Instrum., vol. 77, no. 11, p. 114701 (2006).
\bibitem{RG-RS09}
G. Romero, J. J. Garcia-Ripoll, and E. Solano, ``Photodetection of propagating quantum microwaves in circuit QED", Phys. Scr., vol. T137, p. 014004 (2009).
\bibitem{Cetal11}
Y.-F. Chen {\it et al.}, ``Microwave Photon Counter Based on Josephson Junctions", Phys. Rev. Lett., vol. 107, p. 217401 (2011).
\bibitem{PRJWSG-R11}
B. Peropadre, G. Romero, G. Johansson, C. M. Wilson, E. Solano, and J. J. Garc\'ia-Ripoll, ``Perfect Microwave Photodetection in Circuit QED", Phys. Rev. A, vol. 84, p. 063834 (2011).
\bibitem{SSJ16}
S. R. Sathyamoorthy a, T. M. Stace, and G. Johansson, ``Detecting itinerant single microwave photons", Comptes Rendus Physique, vol. 17,  no. 7, p. 756 (2016).
\bibitem{KKTNN18}
S. Kono, K. Koshino, Y. Tabuchi, A. Noguchi, and Y. Nakamura, ``Quantum non-demolition detection of an itinerant microwave photon", Nature Physics, vol. 14, p. 546 (2018).
\bibitem{Le95}
J. C. G. Lesurf, ``Sky noise". [Online]. Internet: www.st-andrews.ac.uk/\~{}www\_pa/Scots\_Guide/RadCom/part8/ page3.html, 1995 [Accessed: 25- Oct- 2018].
\bibitem{Ca17}
M. Caleffi, ``Optimal Routing for Quantum Networks", IEEE Access, vol. 5, p. 22299 (2017).
\bibitem{CCB18}
M. Caleffi, A. S. Cacciapuoti, and G. Bianchi, ``Quantum Internet:
from Communication to Distributed Computing!", NANOCOM18 Proceedings of the 5th ACM International Conference on Nanoscale Computing and Communication (2018).
\bibitem{GLTM16}
J. Govenius, R. E. Lake, K. Y. Tan, and M. M\"ott\"onen, ``Detection of Zeptojoule Microwave Pulses Using Electrothermal Feedback in Proximity-Induced Josephson Junctions", Phys. Rev. Lett., vol. 117, p. 030802 (2016).
\bibitem{Poetal17}
S. Pogorzalek {\it et al.}, ``Hysteretic Flux Response and Nondegenerate Gain of Flux-Driven Josephson Parametric Amplifiers", Phys. Rev. Applied, vol. 8, p. 024012 (2018).
\end{thebibliography}
\end{document}